\newif\ifpdf
\ifx\pdfoutput\undefined
  \pdffalse              
\else
  \pdfoutput=1           
  \pdftrue
\fi
\documentclass[aps,prb,twocolumn,superscriptaddress]{revtex4}
\ifpdf
  \usepackage[pdftex,
        colorlinks=true,
        pdftitle={Analyzing Single Crystal Time-of-Flight Neutron Data},
        pdfauthor={P.F. Peterson, D.J. Mikkelson, R.L. Mikkelson,
A.J. Schultz, J.P. Hammonds, A. Chatterjee, and T.G. Worlton},
        pdfsubject={},
        pdfkeywords={NOBUGS2002/030},
        pdfpagemode=None,
        bookmarksopen=true]{hyperref}
  \usepackage{thumbpdf}
\fi
\usepackage{graphicx}
\usepackage{amssymb}
\begin{document}

\title{Analyzing Single Crystal Time-of-Flight Neutron Data - NOBUGS2002/030}

\author{P.F. Peterson}
\affiliation{Intense Pulsed Neutron Source, Building 360, 9700 South
Cass Ave, Argonne, IL 60439-4814, USA}
\author{D.J. Mikkelson}
\author{R.L. Mikkelson}
\affiliation{Department of Mathematics, Statistics and Computer
Science, University of Wisconsin-Stout, Menomonie, WI 54751, USA}
\author{A.J. Schultz}
\author{J.P. Hammonds}
\author{A. Chatterjee}
\author{T.G. Worlton}
\affiliation{Intense Pulsed Neutron Source, Building 360, 9700 South
Cass Ave, Argonne, IL 60439-4814, USA}

\date{Oct 19, 2002}

\begin{abstract}
Software for time-of-flight single crystal diffractometer (SCD) data
visualization and analysis was originally written to be run in batch
mode on VMS-VAX systems. Modern computers and software tools allow a
new level of visualization and user interaction. The Integrated
Spectral Analysis Workbench (ISAW) is being extended and customized
for SCD measurements. To this end, new viewers and operators have been
added to ISAW. One of the new operators allows running a program
written in another language, such as C or FORTRAN, to integrate tested
software into the object-oriented package. This provides a method for
preserving earlier software development while making use of modern
tools.

\end{abstract}

\maketitle


\section{Introduction}

The single crystal diffractometer (SCD) at the intense pulsed neutron
source (IPNS) was the among first of its kind when it was built. It
uses a four circle Huber diffractometer with the chi angle fixed at
45~degrees and a single anger camera at 90~degrees from the direct
beam at $\sim$30~cm from the sample. The detector has 85~x~85 pixels
in an active area of 30~cm~x~30~cm.~\cite{schul;taca82} This has been
the instrument configuration, with minor adjustments, ever since. SCD
is currently going through a major upgrade, replacing its original
detector with two higher resolution anger cameras that each have an
active area of 15~cm x 15~cm at a distance of $\sim$20~cm from the
sample. In addition, the data acquisition system (DAS) for the SCD is
also being upgraded. Not surprisingly the entire instrument upgrade is
making it necessary to update much of the analysis software.

The Integrated Spectral Analysis Workbench (ISAW) is a general purpose
data analysis and visualization package which can read the raw data of
any IPNS instrument, including the upgraded SCD. ISAW also has an
extendible system for performing calculations and operations on the
data well suited to developing new analysis methods. This paper will
describe the new data analysis package. Its early versions use much of
the old software while implementing new methods of data visualization.

\section{Current Software}

The intention of all measurements on SCD is to collect up to $2\pi$
steradians of diffraction data from a single crystal sample. The raw
data is then reduced to Bragg intensities consisting of the $(hkl)$
index, integrated intensity, and uncertainty in the integrated
intensity. This information is then used to determine the contents of
the unit cell. To measure the full $2\pi$ steradians up to 44 crystal
orientations need to be measured. For many crystals this is
unnecessary and far fewer orientations are measured to reveal all of
the unique reflections. 
\begin{figure}[h!]
\begin{center}
  \includegraphics[angle=0,width=3.0in]{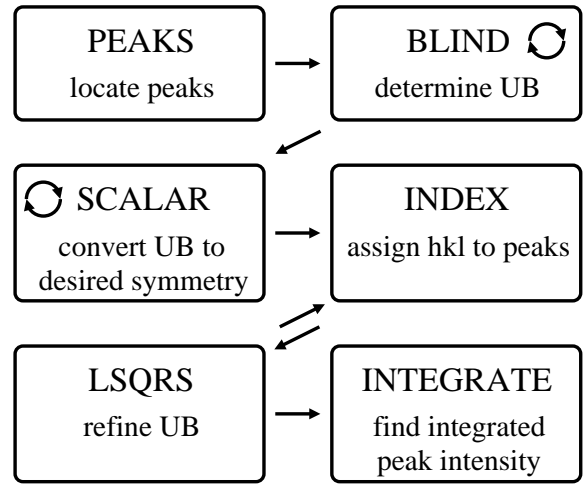}
  \caption{Outline of the data analysis to produce a list of indexed,
  integrated peak intensities.}\label{fig:oldmethod}
\end{center}
\end{figure}
An example of this is quartz, with a hexagonal crystal structure,
which only needs 13 crystal orientations measured to collect all of
the unique reflections.

The original analysis software was written using \mbox{FORTRAN77} to
run in a batch mode on the VAX. The analysis steps are outlined in
Fig.~\ref{fig:oldmethod}. After the analysis is completed the list of
reflections, frequently called peaks, are then converted to a format
appropriate for the modeling software used. {\tt PEAKS} determines the
location of the brightest points in time and pixel space which is
converted to position and wavelength. This is normally run once for
each crystal orientation and all reflections are collected in the same
file. Then {\tt BLIND} is run repetitively with between 10 and 17
reflections to determine lattice parameters ($a$, $b$, $c$, $\alpha$,
$\beta$, $\gamma$) and an orientation matrix ($UB$) which are
satisfactory.~\cite{giaco;bk92,busin;ac67} {\tt SCALAR} transforms the
symmetry of the $UB$ matrix found with {\tt BLIND} to be more
appropriate. For the quartz example mentioned above the $UB$ matrix
determined by {\tt BLIND} will frequently make $b$ the long axis, {\tt
SCALAR} performs the rotation to make $c$ the long axis consistent
with convention. {\tt INDEX} takes the orientation matrix from {\tt
SCALAR} and determines as many of the $(hkl)$ indexes as
possible. Then {\tt LSQRS} uses the indexed peaks to refine the
lattice parameters and the $UB$ matrix, which is then used to run {\tt
INDEX} again to get $(hkl)$ for more peaks. After a few iterations of
{\tt INDEX} and {\tt LSQRS} almost all peaks are indexed and the
lattice parameters and $UB$ matrix are well established. The lattice
parameters and $UB$ matrix are then used one more time to {\tt
INTEGRATE} all possible reflection locations in the data and write a
new peaks file with the integrated intensity.

\section{New Analysis Package}

With the upgraded instrument comes two main concerns: the file format
is being changed and most of the algorithms assume that there is only
one detector. The first problem is dealt with by using ISAW which can
already read the files from the new DAS. The first and last steps in
the analysis, {\tt PEAKS} and {\tt INTEGRATE} respectively, are the
only times that access to the raw data is necessary. For this reason
they are the only ones that need to be modified to deal with the
changing file format. This is accomplished by porting the FORTRAN code
to Java. The other steps in the analysis can be wrapped so ISAW can
have a straightforward method of stepping through the
analysis. Wrapping the software allows for rapid development of the
new analysis package and providing more time for dealing with the more
difficult problem of a second detector. The process of wrapping and
testing software is made significantly simpler through ISAW's operator
mechanism. An operator is similar to a subroutine in most programming
languages. Each operator is intended to be a single step in the
analysis which can be executed in an orderly fashion. For this reason
each of the steps outlined in the previous section becomes an
operator. Then the operators can be linked together with a script in a
very straightforward manner. At the time of publication the migration
to ISAW has been completed through {\tt LSQRS}, and porting {\tt
INTEGRATE} is expected to happen early in 2003.

\section{New Methods of Data Visualization}

\begin{figure}[h!]
\begin{center}
  \includegraphics[angle=0,width=3.0in]{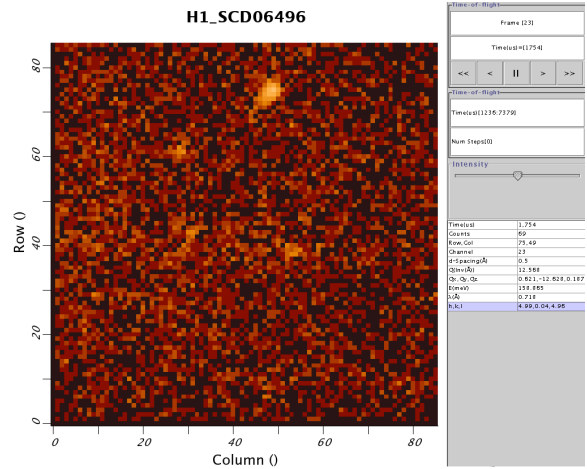}
  \caption{The contour view in ISAW after the orientation matrix,
  $UB$, has been loaded. The selected intensity is the $(550)$ peak of
  the data. The indexes are switched because the refined lattice has
  $b$ as the long axes, contrary to convention.}\label{fig:contour}
\end{center}
\end{figure}

Another feature of ISAW is a set of interactive and highly correlated
viewers. Since the time when the original data visualization software
was written for SCD there has been many improvements in computing. For
this reason the new visualization software can show the data in ways
that were previously not possible. In Fig.~\ref{fig:contour} a
traditional raster plot created by ISAW is shown. To the right of the
plot is a set of readouts that give information about the pixel being
pointed at with the mouse. 
Included in this is the momentum transfer,
$\vec{Q}$, and the index, $(hkl)$, which are both calculated while the
user is interactively exploring the data. Future viewers planed
include:
\begin{itemize}
\item Terrain view which is similar to the raster plot, but with the
intensities plotted as height as well as color.
\item Volume render which produces an overview of the full 3D data by
accumulating and attenuating intensity values along rays through the
data.
\item Slicer that displays a set of three orthogonal intersecting
planes that cut through the data. For example, three planes
$(h_0,k,l)$, $(h,k_0,l)$, and $(h,k,l_0)$ where the values of $h_0$,
$k_0$, and $l_0$ are controlled interactively by the user.
\end{itemize}
Another view in its early development is a splatter plot which
collects all points in $\vec{Q}$-space with an intensity above a
threshold then plots them. This 3D view can be provides zooming and
rotations to better explore the symmetry of the reflections.


\begin{acknowledgments}
PFP would like to thank M. Miller for information concerning the
upgraded SCD configuration. This work was supported by the National
Science Foundation through Grant Number DMR-0218882. Summer support
for this work has benefitted through Department of Educational
Programs and the Summer Faculty Research Program. Argonne National
Laboratory is funded by the U.S. Department of Energy, BES-Materials
Science, under Contract W-31-109-ENG-38.
\end{acknowledgments}


\bibliographystyle{aip}

\end{document}